\documentstyle[prl,aps,epsf]{revtex}
\begin{document}
%\draft
\title{
Anisotropic Fermi Surfaces and Kohn-Luttinger Superconductivity
in Two Dimensions}
\author{J. Gonz\'alez $^1$, F. Guinea $^2$ and M. A. H. Vozmediano $^3$ \\}
\address{
        $^1$Instituto de Estructura de la Materia.
        Consejo Superior de Investigaciones Cient{\'\i}ficas.
        Serrano 123, 28006 Madrid. Spain. \\
        $^2$Instituto de Ciencia de Materiales.
        Consejo Superior de Investigaciones Cient{\'\i}ficas.  
        Cantoblanco. 28049 Madrid. Spain. \\
        $^3$Departamento de Matem\'aticas.
        Universidad Carlos III.
        Avda. de la Universidad 30.
        28911 Legan\'es. Madrid. Spain.}
\date{\today}
\maketitle

\vspace{1cm}
\begin{abstract}
The instabilities induced on a two-dimensional system
of correlated electrons by the anisotropies of its Fermi line are
analyzed on general grounds. Simple scaling arguments allow to
predict the opening of a superconducting gap with a well-defined symmetry
prescribed by the geometry of the Fermi line. The same arguments predict a
critical dimension of 3/2 for the transition of the two-dimensional system 
to non-Fermi liquid behavior. The methods are applied to the $t-t'$
Hubbard model in a wide range of dopings.

\end{abstract}
%\narrowtext %\newpage

\section{Introduction}

Two-dimensional electron systems are the stage in which
most relevant quantum phase transitions arise. The prototype of
these phenomena is the Mott-Hubbard transition\cite{mott},
which can only
be understood in the context of a very strong correlation
between the electrons. Many of the properties of the high-$T_c$
superconductors are also supposed to be the consequence of
strong correlations in the two-dimensional layers\cite{and1}.
In both
cases, these reflect themselves in the presence of very marked
features in the single-particle dispersion relation. The case of
the copper-oxide compounds has been actually studied with
increasing accuracy from the experimental point of view, and the
angle-resolved photoemission measurements have been able to
characterize the drastic change that the dispersion relation
suffers when the stechiometric material is increasingly
doped\cite{photo}.
In the antiferromagnetic insulator regime, the characteristic
peaks at $(\pi/2, \pi/2 )$ and symmetry related points are a
consequence of the doubling of the unit cell and the
hybridization of modes by the wavevector $(\pi, \pi )$\cite{wells}.
In the
superconducting regime, the two-dimensional layers tend to
develop very flat bands at the points $(\pi, 0)$ and
$(0, \pi)$\cite{photo,gofron}.
It is not well-known the mechanism by which the electron system
undergoes such an strong renormalization but, reversing the line
of reasoning, one may ask to what extent the pronounced features
of the Fermi surface are at the origin of the unconventional
properties in the superconducting regime. More generally, there is
a lack of some dynamical formulation predicting the evolution of
the Fermi surface upon doping and, at this point, it is only
possible to characterize the shapes that lead to the opening of
a more stable phase (Mott insulator, superconductor).

In general, phases different from the normal metallic state may
arise from instabilities that have their origin in the peculiar
geometry of the Fermi surface. Such instabilities may lead in
some instances to antiferromagnetism, superconductivity, or the
formation of a charge-density-wave structure. A well-known
example is the case of a Fermi surface with the property of
nesting, that is, with two portions that are mapped one into the
other by some fixed momentum ${\bf Q}$. The system shows then an
enhanced response to a perturbation with such wavevector and, if
the corresponding modulation is commensurate with the lattice,
it may lead to antiferromagnetic order or to a
charge-density-wave structure, depending on the character of the
interaction\cite{shankar}.
On the other hand, there are also instances in
which the geometry of the Fermi surface may give rise to a
superconducting instability, starting from a pure repulsive
interaction. The basis of this mechanism was laid down by Kohn
and Luttinger back in 1965\cite{kl,rusos,rgkl}.
In the case of a 3D electron system
with isotropic Fermi surface, there is an enhanced scattering
at momentum transfer $2 k_F$, which translates into a modulation
of the effective interaction potential $V(r) \sim  \cos (2k_F r)
/r^3 $. This oscillating behavior makes possible the existence
of attractive channels, labelled by the angular momentum quantum
number. The superconducting instability arises from the channel
with the strongest attractive coupling. In two dimensions,
however, it has been shown for the weakly non-ideal Fermi
gas that the effective interaction vertex computed to second
order in perturbation theory
is independent of the
momentum of particles at the Fermi surface\cite{rusos1}.
In this case one has to invoke higher-order effects in the
scattering amplitude in order
to find any attractive channel\cite{chub}.
Otherwise, superconductivity due
to the Kohn-Luttinger mechanism also arises in 2D systems with a
Fermi surface that deviates from perfect isotropy. It has been
studied in the Hubbard model at very low fillings, for instance,
with the result that the dominant instability may correspond 
to $d_{xy}$, $d_{x^2 - y^2}$ or $p$-wave symmetry, depending on
the next-to-nearest neighbor hopping\cite{rusos2,rusos22,hlubina}.

\begin{figure}
\begin{center}
\mbox{\epsfysize 6cm \epsfbox[143 206 440 482]{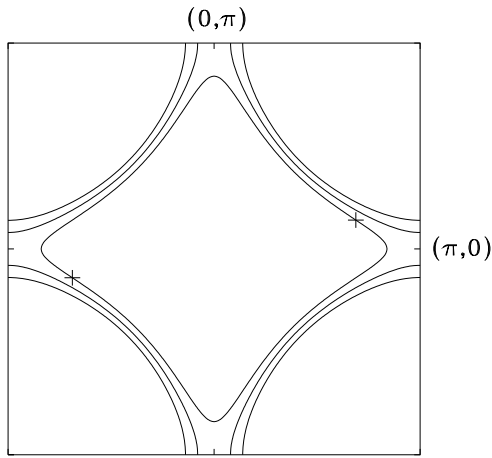}}
\end{center}
\caption{Different shapes of the Fermi line for the $t-t'$ Hubbard
model about the Van Hove filling. Two opposite inflection points are
marked on the figure.}
\label{fermil}
\end{figure}

In the present paper we review the possibility of Kohn-Luttinger
superconductivity in the case of Fermi surfaces with a shape
appropriate to the description of the two-dimensional hole-doped
copper oxide layers. As we want to consider the regime around
the optimum doping for superconductivity, this implies to deal
with highly anisotropic Fermi surfaces. We recall that a most
accurate description of the electronic interaction in the layers
of the hole-doped cuprates is given by the Hubbard model with
one-site repulsive interaction and
significant next-to-nearest-neighbor hopping $t' < 0$\cite{dagotto}.
The values that give the best fit to the Fermi lines determined by
the photoemission experiments seem to be around $-0.3$ times the $t$
hopping parameter. Typical energy contour lines for that value
are shown in Fig. \ref{fermil}. There we can appreciate how the
topology of the Fermi line changes upon doping after passing
through the saddle points at $(\pi, 0)$ and $(0, \pi)$. Quite
remarkably, for filling levels right below the Van Hove
singularity the Fermi line develops inflection points, i.e.
points at which the curvature changes sign. The presence of
these points leads to a strong modulation of the effective
interaction of particles at the Fermi line, as their scattering
is enhanced at momentum transfer connecting every two opposite
inflection points. This effect is reminiscent of what happens in
the case of nesting of the Fermi surface, although in the
present circumstance there are no finite portions of the Fermi
line that are mapped by a given wavevector. As a consequence of
that, there is not a marked tendency towards a magnetic
instability, but the dominant instability corresponds to
superconductivity by effect of the modulation along the
Fermi line of the effective interaction between particles with 
opposite momentum\cite{prl}.

We will see that, in the mentioned cases of strong anisotropy of
the Fermi line, there is a direct correspondence between its
topology and the channel in which the dominant superconducting
instability opens up. As we will be carrying out the discussion
for the $t-t'$ Hubbard model, the channels are characterized by
the irreducible representations of the discrete symmetry group. 
We will show that, in the case in which the Fermi line has a set
of evenly distributed inflection points, superconductivity due
to the Kohn-Luttinger mechanism takes place in the extendend
$s$ channel, with nodes at the inflection points. On the other
hand, when the Fermi line is above the saddle-points, the
superconducting instability most likely takes place in the
$d_{x^2 - y^2}$ channel, with nodes at the points where the
curvature reaches a minimum. Between these two instances, we
find the situation in which the Fermi energy sits at the Van
Hove singularity, what requires special care in order to deal
with the divergences that appear in perturbation theory
associated to the logarithmic density of states. In the context
of highly anisotropic Fermi surfaces, the implementation of the
Kohn-Luttinger mechanism was proposed within the
investigation of the electron system near Van Hove
singularities\cite{epl}.
Related analyses have been also undertaken by other
authors\cite{newns,liu,zanchi}.
As we will see, such an electronic mechanism of
superconductivity near a Van Hove singularity may be relevant to
the physics of the high-$T_c$ 
cuprates\cite{markg,patt,mark}. From the technical point
of view, this is due to the fact that the modulation of the
effective interaction vertex is driven then by a marginal
operator, rather than by an irrelevant operator as it happens in
general, so that in some region of the phase diagram critical
temperatures may be reached as high as those found in the
cuprates\cite{japon}.

We will adopt a renormalization group (RG) approach to the discussion
of the subject. In recent years, RG methods have been applied to
the description of interacting electron systems\cite{shankar,pol}.
They have given
a precise characterization of Fermi liquid theory, as well as an
efficient classification of its relevant perturbations\cite{shankar}.
Among a
reduced number of them, superconductivity appears as one of the
possible instabilities. It only requires the presence of an
attractive coupling in some channel, though small it may be in
the bare theory. RG methods are also particularly well-suited to
the discussion of highly anisotropic Fermi surfaces, since a
change of the topology is accompanied by a change of the scaling
dimension of the interaction vertex. This is the way in which
the enhancement of scattering produced by some features, as for
instance the inflection points, is understood in the RG
framework\cite{prl}.
The RG approach is specially useful when dealing with
unconventional mechanisms of superconductivity, as it may face
the possibility that the normal state does not fall into the
Fermi liquid description. Under this circumstance, it may not be
justified to use a perturbative approach, or even to rely on 
certain sum of diagrams in perturbation theory, to study the
superconducting instability. Yet in the RG approach the only
basic assumption is that there is a well-defined scaling near
the Fermi level, which allows to identify the scaling operators
in the low-energy effective theory. Apart from addressing the
question of superconductivity, the discussion can be devoted
to establish when such scaling
conveys to non-Fermi
liquid behavior\cite{wen}, spoiling the conventional picture that
considers superconductivity as a low-energy instability of Fermi
liquid theory.

The paper is organized as follows. In the next section
we briefly review the renormalization group approach to
interacting electron systems.
In section 3 we investigate under which conditions non-Fermi
liquid behavior may arise by effect of the anisotropy of the
Fermi surface. 
The analysis of the
superconducting instabilities in the presence of inflection
points in the Fermi line is carried out in section 4. 
The RG approach is adapted to the discussion of the 2D electron
system near Van Hove singularities in section 5.
Finally, our conclusions and outlook are drawn in
section 6.

\section{Renormalization Group Approach to Interacting Electrons}

The wilsonian RG approach, that has proven to be so useful
in the study of classical statistical systems, has been recently
implemented in the investigation of many-body systems with a
Fermi surface\cite{shankar}. 
In this context, the basic idea of the method is
closely related to the concept of effective field theory\cite{pol}. 
One is
interested in identifying the elementary fields and excitations
at a very low energy scale about the Fermi level, that is, at a 
much smaller scale than the typical electron energies of the
bare theory. The strategy to accomplish this task is to perform
a progressive integration of high-energy modes living in two
thin shells at distance $\Lambda $ in energy below and above the
Fermi surface. Sufficiently close to it, the fields of the
effective theory have to scale appropriately under a reduction
of the cutoff $\Lambda \rightarrow  s \Lambda $. Then, one can
make an inspection of all the possible terms built out of them
contributing to the effective action of the theory, checking
their behavior under the scale transformation
\begin{equation}
g_i \; \int dt d^D p \; {\cal O}_i \rightarrow
              g_i \; s^{\alpha_i } \;  \int dt d^D p \; {\cal O}_i
\end{equation}

The terms in the effective action that scale with an exponent
$\alpha_i > 0$ are said to be irrelevant, as their effect
becomes weaker and weaker close to the Fermi surface. On the
other hand, if there appear some terms scaling with $\alpha_i <
0$, this means that we have chosen a wrong starting point for
the effective field theory, since there are couplings that are
not stabilized already at the classical level. The interesting
case corresponds to having all the exponents $\alpha_i \geq 0$
and, in particular, operators with $\alpha_i = 0$ (so called
marginal operators),
as this is the situation where the RG approach can address the
existence of a fixed-point of the scale transformations in the
low-energy theory. 

\begin{figure}
\begin{center}
\mbox{\epsfysize 6cm \epsfbox[176 346 449 526]{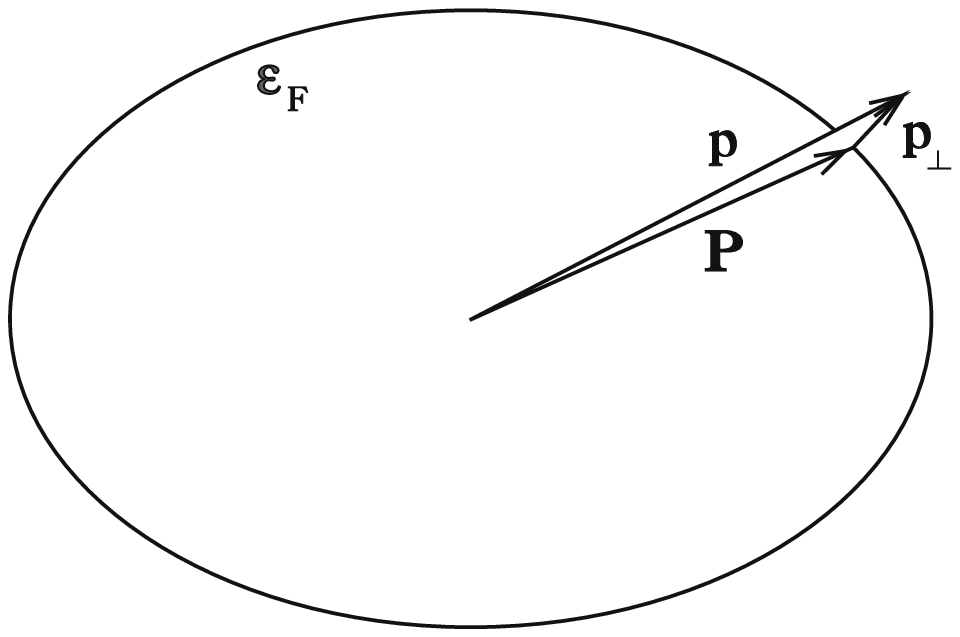}}
\end{center}
\caption{Decomposition of the momentum ${\bf p}$ into a transverse
and a large component at the Fermi line.}
\label{decomp}
\end{figure}

The system of interacting electrons with a sphere-like
Fermi surface provides a good example of how the above program
is at work\cite{follow}. 
We focus from now on in two spatial dimensions and
write the action with the most general four-fermion interaction
term 
\begin{eqnarray}
S  & = & \int dt d^2 p \left( i \Psi^+_\sigma ({\bf p}) \partial_t
   \Psi_\sigma ({\bf p}) - \left( \varepsilon ({\bf p}) - \varepsilon_F
  \right) \Psi^+_\sigma ({\bf p}) \Psi_\sigma ({\bf p})  \right)  
                                     \nonumber  \\
 &  &  + \int dt d^2 p_1 d^2 p_2 d^2 p_3 d^2 p_4    \;
   U ({\bf p}_1, {\bf p}_2, {\bf p}_3, {\bf p}_4)   \;
      \Psi^+_\sigma ({\bf p}_1)
   \Psi^+_{\sigma'} ({\bf p}_2) \Psi_{\sigma'} ({\bf p}_4) 
       \Psi_\sigma ({\bf p}_3) 
  \delta ( {\bf p}_1 + {\bf p}_2 - {\bf p}_3 - {\bf p}_4 )
\label{effact}
\end{eqnarray}
We suppose that the integration of high-energy modes has already
proceed to energies close to the Fermi level, so that $\Psi
({\bf p})$ stands here for some renormalized scaling fermion
field. The important point to notice is that, at each point of
the Fermi line, only the component of the momentum orthogonal to
it scales with the energy cutoff. In fact, near the Fermi level
we can decompose every momentum ${\bf p}$
into a vector ${\bf P}$ to the closest point in the Fermi line
and the orthogonal component ${\bf p}_\perp $, as shown in Fig.
\ref{decomp},
\begin{equation}
{\bf p} = {\bf P} + {\bf p}_\perp
\end{equation}
Upon the scale reduction of the energy cutoff, we have ${\bf p}_\perp 
\rightarrow s {\bf p}_\perp $.

It becomes clear, for instance, that in the noninteracting
theory the field $\Psi ({\bf p})$ has a well-defined scaling
rule that makes the first term of (\ref{effact}) scale invariant. 
We can rewrite the free effective action in the form 
\begin{equation}
S_0   \sim  \int dt d P d p_\perp 
    \left( i \Psi^+_\sigma ({\bf p}) \partial_t
   \Psi_\sigma ({\bf p}) - v_F ({\bf P}) p_\perp 
        \Psi^+_\sigma ({\bf p}) \Psi_\sigma ({\bf p})  \right)
\label{free}
\end{equation}
Under a change of the cutoff $\Lambda \rightarrow s \Lambda $
and corresponding changes in the scaling variables $t
\rightarrow s^{-1} t, p_\perp \rightarrow s p_\perp $, the field
has to transform according to $\Psi ({\bf p}) \rightarrow
s^{-1/2}  \Psi ({\bf p})$ in order to keep (\ref{free}) marginal.

A similar analysis applied to the interaction term of the
effective action
\begin{eqnarray}
S_{int} & \sim & \int dt d P_1 d p_{1\perp} d P_2 d p_{2\perp}
     d P_3 d p_{3\perp}  d P_4 d p_{4\perp}   \nonumber \\
 &  &   U ({\bf p}_1, {\bf p}_2, {\bf p}_3, {\bf p}_4) \;
    \Psi^+_\sigma ({\bf p}_1)
   \Psi^+_{\sigma'} ({\bf p}_2) \Psi_{\sigma'} ({\bf p}_4) 
        \Psi_\sigma ({\bf p}_3) 
  \delta ( {\bf p}_1 + {\bf p}_2 - {\bf p}_3 - {\bf p}_4 )
\end{eqnarray}
leads to the conclusion that, in general, it scales in the form
$S_{int} \rightarrow s S_{int} $. We are assuming that, upon
scaling, the orthogonal component of the momenta are irrelevant
in the momentum-conservation delta function, compared to the
large ${\bf P}$ components, so that $\delta ( {\bf p}_1 + {\bf
p}_2 - {\bf p}_3 - {\bf p}_4 ) \approx \delta ( {\bf P}_1 + {\bf
P}_2 - {\bf P}_3 - {\bf P}_4 ) $. Thus the interaction turns out
to be irrelevant for generic values of the momenta (assuming
implicitly a smooth dependence on them of the potential 
$U ({\bf p}_1, {\bf p}_2, {\bf p}_3, {\bf p}_4)$).

The important remark put forward in Refs. \onlinecite{shankar} and
\onlinecite{pol} is that there are special processes in which
the kinematics forces a different scaling behavior through the
constraint of momentum conservation. This is the case when the
combination of the momenta at the Fermi surface identically
vanishes, ${\bf P}_1 + {\bf P}_2 - {\bf P}_3 - {\bf P}_4 = 0$.
When this happens, the scaling of the orthogonal components of
the momenta is transferred to the scaling of the delta function,
and the four-fermion interaction term becomes marginal. This
explains why in Fermi liquid theory there are only a few
channels that are not irrelevant 
in the low-energy theory.

\begin{figure}
\begin{center}
\mbox{\epsfysize 3cm \epsfbox[76 595 521 695]{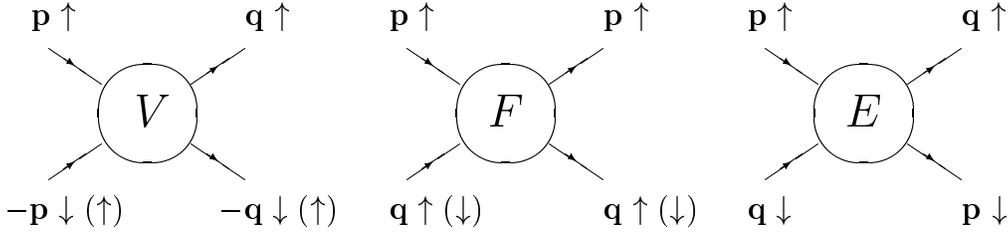}}
\end{center}
\caption{Marginal interactions in Fermi liquid theory.}
\label{vfe}
\end{figure}

Specifically, the above identity for the sum of the components
at the Fermi surface is satisfied by i) ${\bf P}_1 = - {\bf P}_2$
and ${\bf P}_3 = - {\bf P}_4$, what characterizes the so called
BCS channel, ii) ${\bf P}_1 =  {\bf P}_3$ and ${\bf P}_2 = {\bf
P}_4$, that leads to the forward scattering channel, and iii)
${\bf P}_1 =  {\bf P}_4 $ and ${\bf P}_2 =  {\bf P}_3$, that
differs from the previous one by the exchange of the outgoing
particles. In this classification we have not taken into account
the spin of the particles, but it is clear that, if in the third
case the spin of ${\bf P}_4$ is different to that of ${\bf
P}_1$, the scattering process is fully differenciated from that
of forward scattering. In the case of a spin-dependent interaction, 
it makes sense to
consider iii) as a different channel on its own, that we will
call exchange scattering channel\cite{mcc}. 
The three different channels
leading to marginal interactions are represented graphically in
Fig. \ref{vfe}.

We address now the behavior of the marginal interactions under
quantum corrections by focusing again on a system with a Fermi line
having the topology of the circle. This is actually the
case best studied in the literature. We borrow the main
conclusions from Refs. \onlinecite{shankar} and \onlinecite{pol},
which state, to begin with, that the interaction in the forward
scattering channel remains unrenormalized in the quantum theory.
That is, the integration of the high-energy modes does not
produce any dependence of such marginal interaction on the
cutoff by effect of the loop corrections. This is the way in
which the Landau theory of the Fermi liquid is recovered in the
present context, with a {\em finite} correction of the bare
parameters and a $F$ function that encodes all the information
about the four-fermion interaction.

On the other hand, the pairing interaction in the BCS channel
gets dependence on the cutoff by integration of modes in the two
thin slices at energies $\Lambda $ and $-\Lambda $. At the
one-loop level, for instance, the particle-particle bubble is
the diagram that contributes to the renormalization since, if
one of the particles in the loop carries momentum ${\bf p}$ at 
the slice being integrated, the other has a momentum $-{\bf p}$ 
that also belongs to the slice. In two dimensions, taking into
account the dependence of the BCS coupling on the angles $\theta
$ and $\theta '$ of the respective incoming and outgoing pairs,
the one-loop differential correction becomes
\begin{equation}
V_{\Lambda + d\Lambda} (\theta, \theta ')  \approx
 V_{\Lambda } (\theta, \theta ') +  N(\varepsilon_F )
   \; d\Lambda \;  \int d\theta '' V_{\Lambda } (\theta, \theta '')
     V_{\Lambda } (\theta '', \theta ') \;\;\;\; , \;\;\;\;
            d\Lambda < 0
\end{equation}
where $N(\varepsilon_F)$ is the density of states at the Fermi
level. 

The one-loop flow equation
\begin{equation}
\Lambda \frac{\partial V (\theta, \theta ')}{\partial \Lambda } 
 = N(\varepsilon_F ) \int d\theta '' V (\theta, \theta '')
     V (\theta '', \theta ')
\end{equation}
describes the approach to the Fermi line as $\Lambda \rightarrow
0$. Thus, in systems where the original bare interaction is
repulsive the BCS channel has a general tendency to become
suppressed at low energies. It is only when $V$ or any of its
normal modes is negative that the renormalized coupling becomes
increasingly attractive near the Fermi level. 
This is the way in which the superconducting instability arises 
in the RG framework\cite{shankar,pol}.
The main conclusion is that the system with uniform repulsive
interaction and isotropic Fermi line provides a paradigm
of the Landau theory of the Fermi liquid, since all the
interactions turn out to be irrelevant but the $F$ interaction,
that remains finite and non-zero  close to the Fermi
level.

The above RG scheme gives also the clue of why the slight
deviation from isotropy may trigger the onset of
superconductivity. In general the analysis has to be carried
over the couplings $V_n$ for the normal modes appropriate to the
symmetry of the Fermi line. The flow equation may be decomposed
then into
\begin{equation}
\Lambda \frac{\partial V_n}{\partial \Lambda} = N(\varepsilon_F)
 V^2_n
\end{equation}
and the flow of the couplings is given by
\begin{equation}
V_n (\Lambda ) = \frac{V_n (\Lambda_0 )}{1 + N(\varepsilon_F)
 \; V_n (\Lambda_0 ) \;  \log(\Lambda_0 / \Lambda )}
\end{equation}
If, at some intermediate scale $\Lambda_0$, a coupling
$V_n (\Lambda_0 )$ becomes
negative, then a superconducting instability develops in the
corresponding mode. The Kohn-Luttinger mechanism is based on the
fact that this happens in practice, even for an isotropic Fermi
surface in three dimensions, since the corrections to the bare
$V(\theta, \theta ')$ by particle-hole diagrams already
introduce the degree of anisotropy needed to turn some of the
$V_n$ negative\cite{kl}. 
In two dimensions, the anisotropy of
the Fermi line is also a necessary condition to produce an
attractive interaction in some of the normal modes, as pointed out
in Ref. \onlinecite{rusos1}, unless the particle-hole corrections
to the bare $V(\theta, \theta ')$ are carried out beyond second order in
perturbation theory\cite{chub}.

Previous analyses of the Kohn-Luttinger mechanism have focused
on small deviations from isotropy of the Fermi line\cite{rusos}. 
The
relevance of the superconducting transition is related anyhow
to the source of anisotropy that is producing the 
attractive coupling in the model. An attractive but very tiny
$V_n (\Lambda_0 )$ may lead to superconductivity at a
very small scale $E \sim \Lambda_0 
\exp \{1/ \left( N(\varepsilon_F)
 V_n (\Lambda_0 ) \right) \}$. Our RG analysis makes clear
that the anisotropy has an effect through corrections to the bare 
scattering $V(\theta, \theta ')$ that are irrelevant as the
cutoff $\Lambda \rightarrow 0$. We will see that they scale in
general like $\Lambda^{1/2} $, that is, with the same scaling law 
as for scattering with $2k_F$ momentum transfer in the isotropic
Fermi line. On the other hand, we will be dealing with topological 
features of the Fermi line, i.e. inflection points, that enhance
the attractive modes to scale like $\Lambda^{1/4} $\cite{prl}.
Inflection points can be thought of as precursors of the
development of the saddle points at the Fermi line, a situation
in which the particle-hole corrections to scattering processes
become marginal ($\sim \Lambda^0$ ). Consequently, in the context 
we are facing of a repulsive bare interaction,
the superconducting instability is significantly
strengthened near the Van Hove 
singularity\cite{epl,newns,liu,zanchi,rice}, 
although then there is also competition with magnetic
instabilities in the system\cite{mark,schulz,lederer,dz,japon}.
The RG analysis of the problem becomes quite delicate, moreover,
as one has to derive a proper scaling in a model which has enhanced
$\log^2 (\Lambda )$ divergences in the BCS channel. 
The special topology
of the Fermi line passing by the saddle points introduces significant
changes in the analysis carried out for the Fermi liquid
theory and it requires the specific treatment 
that we will discuss in Section 5.

\section{Inflection points of the Fermi line}

As discussed in the preceding section, the scaling of the different
couplings can be read off from the cutoff dependence of the 
particle-particle and particle-hole diagrams which modify them. When the
cutoff
is sufficiently close to the Fermi surface, these diagrams are reduced
to convolutions of single particle propagators taken at the Fermi
surface. The relevance or irrelavance of the couplings is determined
by the density of states of electron hole pairs, or of two electron
states.

We can deduce the real part of a given diagram from its
imaginary part by performing a Hilbert transform. The
imaginary parts measure directly the density of excited states,
and are usually easier to estimate. Let us now imagine that
we can determine that, for a given susceptibility, $\chi$,
the low energy behavior is ${\rm Im} \chi ( {\bf q} , \omega )
\sim \omega^{\alpha}$. Then, ${\rm Re} \chi ( {\bf q} , \omega )$
includes a term with the same $\omega$ dependence if
$\alpha \ne 0$, and shows a logarithmic behavior if
$\alpha = 0$. This contribution will dominate the
corrections to the couplings when it is present.

The quantity ${\rm Im} \chi ( {\bf q} , \omega )$ is a measure of
the electron-hole or electron-electron excitations of total
momentum ${\bf q}$ and energy $\omega$. The simplest way to
estimate it is to measure the total number of excitations
with momentum ${\bf q}$ and energy equal or less than $\omega$.
By deriving the volume of phase space which fulfills these
contraints with respect to $\omega$, one obtains ${\rm Im}
\chi ( {\bf q} , \omega )$.

The leading  diagram which has a non trivial
low energy behavior is the electron electron propagator at 
zero momentum, the BCS channel, as discussed in the
previous section. It is easy to show that ${\rm Im} \chi_{BCS} ( {\bf q}
\rightarrow 0 , \omega )
\rightarrow$ constant as $\omega \rightarrow 0$.
This diagram describes the main instability of Fermi liquids.
In the following, we analyze the modifications that
insertions of electron-hole propagators induce in this
channel, following the ideas of Kohn and Luttinger.

\begin{figure}
%\begin{picture}(235,225) (-0,20)
%\epsfbox{fig1:inflection.eps}
%\end{picture}
%\epsfxsize=\hsize %\epsfysize = 4cm
%\centerline{\epsfbox{fig1:inflection.eps}} \vspace{-1.5cm}

\begin{center}
\mbox{\epsfxsize 5cm \epsfbox{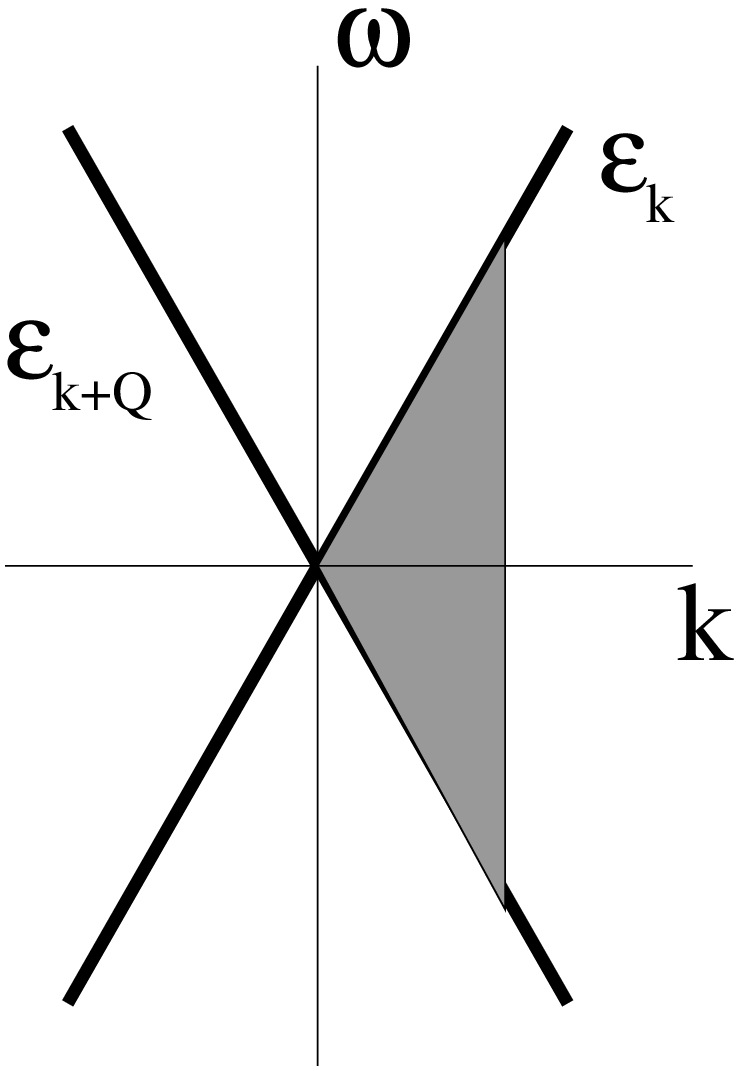}}
\end{center}
\caption{Available phase space for particle hole excitations
with momentum $Q = 2 k_F$ and energy less or equal to $\omega$.
One of the two branches of the electronic band has been displaced
by $Q$.}
\label{fig1_inflection}
\end{figure}

We first need to estimate the the energy dependence of
the electron hole propagator ${\rm Im} \chi ( {\bf q} , \omega )$.
In one dimension, it is relatively straightforward to show that the
imaginary part of the electron-hole susceptibility, 
${\rm Im} \chi ( q , \omega )$, is finite all the way to zero
energy only when $q = 2 k_F$. 
The limit of ${\rm Im} \chi ( 2 k_F , \omega )$ as $\omega 
\rightarrow 0$ goes as the single particle density
of states at the Fermi level (a schematic view of the available
density of states is shown in Fig. \ref{fig1_inflection} ).  

From $\lim_{\omega \rightarrow 0} {\rm Im} \chi ( 2 k_F , \omega )
\propto
{\rm Im} G ( \varepsilon_F )$, we deduce that $\lim_{\omega \rightarrow 0}
{\rm Re} \chi ( 2 k_F , \omega ) \propto {\rm Im} G ( \varepsilon_F )
\log ( \Lambda / \omega )$, where $\Lambda$ is the cutoff of the theory.
Outside $q = 2 k_F$ there are no gapless electron-hole excitations in
one dimension. The log dependence in ${\rm Re} \chi$, when inserted in
the appropiate diagrams, leads to the existence of marginal couplings,
and to the non trivial phenomenology of Luttinger liquids.

\begin{figure}
\begin{center}
\mbox{\epsfxsize 5cm \epsfbox{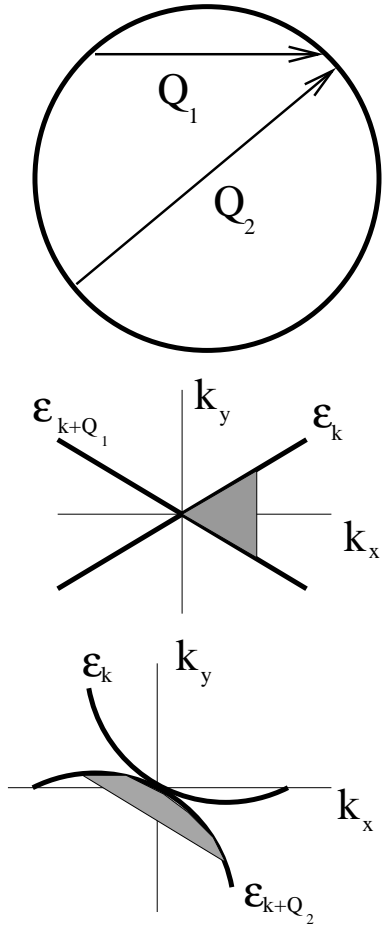}}
\end{center}
\caption{
Circular Fermi surface (top) and the two possible behaviors
of ${\rm Im} \chi ( {\bf Q} , \omega )$ at low energies.
Middle: Generic case of non parallel patches of the  
Fermi surface. Shaded area is the available phase space.
Bottom: Wavevector $Q_2$ which connects two regions where
the Fermi surface is parallel in the two patches.}
\label{fig2_inflection}
\end{figure}

In higher dimensions, ${\rm Im} \chi ( {\bf q} , \omega )$ remains finite
at low energies when   
${\bf q}$ connects two points at the Fermi surface.
The density of electron-hole pairs at this wavevector
depends on the curvature of the Fermi surface at these two points,
and also on their relative orientation. In general, the two patches of the
Fermi surface are not parallel. We can linearize the dispersion   
relation as $\varepsilon ({\bf k}_i+ {\bf k}') = {\bf v}_F^i {\bf k}'$, where
${\bf k}_1 = {\bf k}_0$ and ${\bf k}_2 = {\bf k}_0 + {\bf q}$ denote
two points at the Fermi surface. The constraints described earlier
imply that:
$\omega \ge \varepsilon ({\bf k}_1 + {\bf k}') -
\varepsilon ({\bf k}_2 + {\bf k}')$, and, simultaneously,
$\varepsilon ({\bf k}_1 + {\bf k}') > 0$ and
$\varepsilon ({\bf k}_2 + {\bf k}') < 0$.  
These three conditions, if ${\bf v}_F^1$
and ${\bf v}_F^2$ are not parallel,
define a triangle in ${\bf k}'$ space such that the
basis and the height are proportional to $\omega$,
as schematically shown in Fig. \ref{fig2_inflection}.
Thus, ${\rm Im} \chi ( {\bf q} , \omega ) \propto \omega$.
This linear dependence on $\omega$ remains when the local
density of electron-hole excitations is determined, by
integrating over ${\bf q}$. This is the relevant   
quantity which determines the coupling of impurities to the
electron liquid, and can lead to non trivial
divergences in perturbation theory, as best shown in
the Kondo model.

The previous analysis does not hold when ${\bf q}$ connects two
regions of the Brillouin Zone where the Fermi surface is parallel.
The simplest case is also shown in Fig. \ref{fig2_inflection}.
The linear expansion of the dispersion relation is not enough to
estimate ${\rm Im} \chi$. Using a frame of reference such that one
axis is perpendicular to the two Fermi surfaces, one can write:
\begin{eqnarray}
\varepsilon ({\bf k}_1 + {\bf k}') &= &v_F^1 k'_{\perp} +   
\alpha_1 k'^2_{\parallel}    \nonumber        \\
\varepsilon ({\bf k}_2 + {\bf k}') &= &v_F^2 k'_{\perp} +
\alpha_2 k'^2_{\parallel}
\label{quadratic}
\end{eqnarray}
The boundaries in ${\bf k}'$ of the region where particle-hole
excitations of energy less than $\omega$
are possible are given by $k'_{\perp} \sim
\omega / ( v_F^1 + v_F^2 )$ and $k'_{\parallel} \sim
\sqrt{\omega / ( \alpha_1 + \alpha_2 )}$, giving a total
area $\propto \omega^{3/2}$. Hence,
${\rm Im} \chi ( {\bf q} , \omega ) \sim \sqrt{\omega}$.
This analysis is restricted to two spatial dimensions.   
In higher dimensions, one has to take into account the
additional $D-2$ dimensions. Along each of them, the
available area is limited to a vector in $k$ space which
also scales as $\sqrt{\omega}$. Hence, the volume is
$\propto \omega^{(D+1)/2}$, and ${\rm Im} \chi (
{\bf q}, \omega ) \sim \omega^{(D-1)/2}$. There is
a logarithmic divergence in ${\rm Re} \chi$ when $D = 1$,
which plays the role of the upper critical dimension.

\begin{figure}
\begin{center}
\mbox{\epsfxsize 5cm \epsfbox{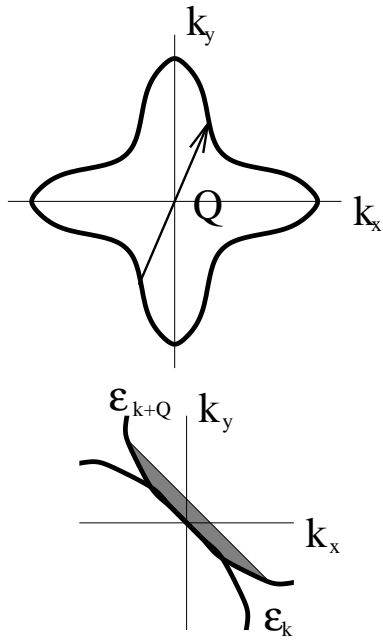}}
\end{center}
\caption{
Top: Anisotropic Fermi surface with inflection points.
Bottom: Available phase space when $Q$ connects two
inflection points.}
\label{fig3_inflection}
\end{figure}

We now assume that the Fermi surface is sufficiently 
anisotropic, as shown in Fig. \ref{fig3_inflection},
so that there are inflections points
at wavevectors $\{ {\bf k}_i \}$. A momentum transfer
equal to ${\bf q}_i = 2 {\bf k}_i$ connects the two
inflection points at $\pm {\bf k}_i$ which, in addition,
define parallel patches of the Fermi surface. The expansion
in Eq. (\ref{quadratic}) is not enough, as $\alpha_1 =   
\alpha_2 = 0$. It must be replaced by:
\begin{eqnarray}   
\varepsilon ({\bf k}_1 + {\bf k}') &= &v_F k'_{\perp} +
\beta k'^3_{\parallel} + \gamma k'^4_{\parallel}  \nonumber \\
\varepsilon ({\bf k}_2 + {\bf k}') &= &- v_F k'_{\perp} -
\beta k'^3_{\parallel}  + \gamma k'^4_{\parallel}    
\label{quartic}
\end{eqnarray}
where we are using the symmetry between the two points.
The area with electron-hole pairs with energy equal or less  
than $\omega$ is now limited by $k'_{\perp} \sim \omega
/ v_F$ and $k'_{\parallel} \sim ( \omega / \gamma )^{1/4}$. 
Hence, ${\rm Im} \chi ( {\bf q} , \omega ) \sim
\omega^{1/4}$.

The previous argument can be extended to arbitrary dimensions.
In general, an inflection point along one of the $D - 1$ transverse
directions at the Fermi surface will not coincide with an
inflection point along another direction. Thus, in the
remaining $D - 2$ directions, the phase space is limited
by quadratic terms, obtained from an expansion like that
in Eq. (\ref{quadratic}). Hence, ${\rm Im} \chi ( {\bf q} ,
\omega ) \sim \omega^{(D - 2) /2 + 1 / 4}$. The dependence of
${\rm Re} \chi$ on the high energy cutoff of the theory
becomes logarithmic when $D = 3 / 2$ \cite{prl}. In the RG
language, the couplings at wavevector ${\bf q}$ are
marginal, and require a non trivial renormalization, at this dimension.
It is interesting to note that, for isotropic Fermi surfaces,
the critical dimension is 1\cite{UR84,CCM94}.

\section{Anisotropic Fermi lines and Kohn-Luttinger superconductivity}

In the previous Section we have seen that the strength of electron
scattering bears a direct relation with the curvature at each
point of the Fermi line. When this becomes highly anisotropic,
the strong modulation in the angular
dependence of the scattering between particles with momentum
${\bf p}$ and ${\bf -p}$ may lead to a superconducting instability,
as explained in Section 2. We will address here the effect of
inflection points in the $t-t'$ Hubbard model,
that is the simplest model with this kind of features in the 
Fermi line, and we will determine the symmetry
channels in which pairing can take place.

The main issue is whether there exists any attractive coupling,
at some intermediate energy scale $\Lambda_0$, for some of the
normal modes of the bare BCS vertex $V(\Lambda_0)$. Here
$\Lambda_0$ is supposed to be a small fraction of the whole
bandwidth, but not yet at the final stage of the renormalization
process. We recall that $V$ depends on the angles $\theta $ and
$\theta '$ of the respective incoming and outgoing pairs of  
electrons, so that it may be decomposed 
into eigenfunctions with well-defined transformation properties
under the action of the lattice symmetry group $D_4$. This has 
four one-dimensional representations, labelled respectively
by $A_1, A_2, B_1$ and $B_2$, and a two-dimensional 
representation labelled by $E$. The complete sets of eigenfunctions 
for the respective irreducible representations are given by
\begin{eqnarray}
A_1   &  :  &  \left\{ \cos (4n \theta ) , n \in \rm{\bf N}  \right\} 
                             \nonumber       \\
A_2   &  :  &  \left\{ \sin (4n \theta ) , n \in \rm{\bf N}  \right\}  
                           \nonumber              \\ 
B_1   &  :  &  \left\{ \cos \left( (4n + 2 )\theta \right) , n \in \rm{\bf N}  
                               \right\}  \nonumber       \\ 
B_2   &  :  &  \left\{ \sin \left( (4n + 2 )\theta \right) , n \in \rm{\bf N}
                               \right\}   \nonumber      \\
E   &  :  &  \left\{ a \sin \left( (2n + 1 )\theta \right) 
             +  b \cos \left( (2n + 1 )\theta \right), n \in \rm{\bf N}
                               \right\}  
\end{eqnarray}

Thus, the BCS vertex may be expanded in the form
\begin{eqnarray}
V( \theta , \theta ')  & = & V_0 + V_1 \left( \cos (\theta ) \cos
(\theta ') +  \sin (\theta ) \sin (\theta ') \right) +
V_2  \cos (2\theta ) \cos (2\theta ') +  V_3 \sin (2\theta ) 
                              \sin (2\theta ')  \nonumber   \\
  &  & + V_4 \left( \cos (3\theta ) \cos
(3\theta ') +  \sin (3\theta ) \sin (3\theta ') \right)  
+ V_5 \cos (4\theta ) \cos (4\theta ') +  V_6 \sin (4\theta ) 
                        \sin (4\theta ')  +  \ldots  
\label{exp}
\end{eqnarray}
According to the discussion in Section 2, it suffices that any
of the $V_n$ couplings becomes negative for a superconducting
instability to take place in the process of renormalization,
with the corresponding symmetry of the order parameter
---$s$-wave in the case of $A_1$ and $A_2$,
and $d$-wave in the case of the $B_1$ and $B_2$
representations. 

We know that all the interactions are irrelevant in Fermi liquid
theory, except the marginal $V$, $F$ and $E$ interactions.
Furthermore, a repulsive interaction in the $V$ channel flows to
zero upon renormalization towards the Fermi level. The point is
that, even starting with a bare repulsive interaction as in the 
$t-t'$ Hubbard model, some of the irrelevant couplings may lead
to attraction in the BCS channel well before arriving at the
final stage of the renormalization process. Under these
circumstances, one needs to make an inspection of the theory at
the intermediate scale $\Lambda_0$, in which the irrelevant
interactions discussed in Section 2 have not become negligible
yet. They may turn negative, in fact, some of the $V_n$
couplings. When this happens, by considering $\Lambda_0$ as
the starting point of the renormalization group flow one
observes the appearance of a strong attraction that leads to
superconductivity in the low-energy effective theory.

\begin{figure}
\begin{center}
\mbox{\epsfysize 6cm \epsfbox[90 369 336 482]{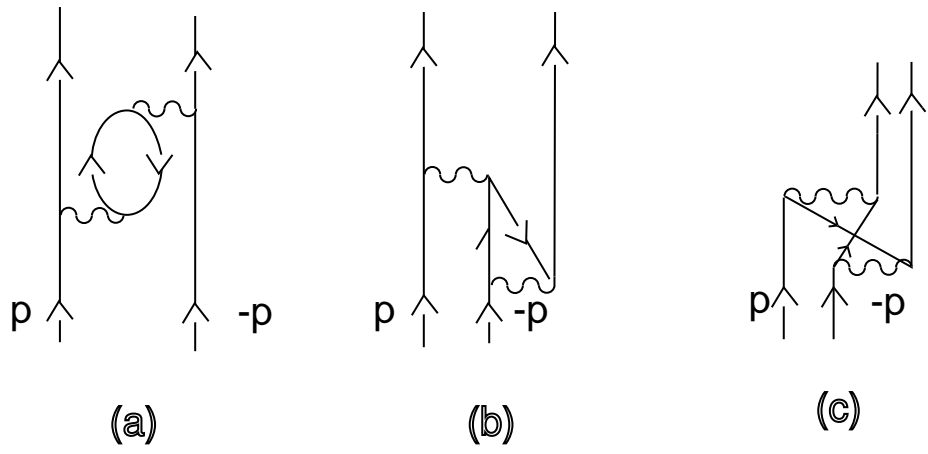}}
\end{center}
\caption{Diagrams building up the effective interaction in the
BCS channel.}
\label{prl}
\end{figure}

Before switching on the quantum corrections, the marginal
interaction $V$ equals the bare on-site repulsion $U$ of the $t-t'$
Hubbard model. Given that the corrections we want to study at
the intermediate scale $\Lambda_0$ are given by irrelevant
operators, we may estimate their effect in perturbation theory. 
At the one-loop order, the diagrams that contribute in general
are those in Fig. \ref{prl}. In the case of the local Hubbard
interaction, we take the convention of writing it as mediated by
a potential between currents of opposite spin. This makes clear
that the first two diagrams in Fig. \ref{prl} cannot be drawn
with the interaction at hand, since they require an interaction
between currents with parallel spins. They are present in the
case of more general short-range interactions, but they tend to
cancel out when such range shrinks to zero.

The contribution 
$\Pi (\theta , \theta ')$ of  diagram $(c)$ 
has the overall effect of reinforcing
the bare repulsion between the electrons. 
We recall that $\theta $ and $\theta' $ label  
the angles between the incoming momentum ${\bf p}$ and the
outgoing momentum ${\bf k}$, respectively, and the $x$ axis.
When the two angles are the same, the momentum flowing in the
interactions is large and connects opposite points at the Fermi
line. We may borrow then the results of the previous Section to
determine the scaling of the particle-hole contribution
$\Pi (\theta , \theta )$ correcting 
$V( \theta , \theta )$.

The simplest case to discuss is that of a set of eight inflection
points on the Fermi line evenly distributed in the angular
variable $\theta $. Then the function $\Pi (\theta , \theta )$
must admit an approximate representation of the form 
\begin{equation}
\Pi (\theta , \theta ) \approx a + b \sin^2 (4 \theta )
\end{equation}
We know from the previous Section that the coefficients $a$ and
$b$ have different order of magnitude, given their different
dependence on the energy cutoff $\Lambda_0 $. At $\theta = 0$,
the scattering does not differ much from that of momentum
transfer $2 k_F$ on an isotropic Fermi line, so that 
$a \sim O ( \Lambda_0^{1/2} / \alpha^{1/2} )$, with the notation
of the previous Section. 
At $\theta = \pi /8$,
the scattering is greatly enhanced since it takes place between
particles sitting at opposite inflection points on the Fermi
line, implying $b \sim O ( \Lambda_0^{1/4} / \gamma^{1/4} )$.

The approximate expression for the bare BCS vertex
at the energy scale $\Lambda_0$ 
\begin{equation}
U + \Pi ( \theta , \theta ) \approx  U + a + b \sin^2 (4\theta )
\end{equation}
matches well the expansion (\ref{exp}) truncated to include up
to the $V_6$ term. This allows to find the scaling of some
combinations of the coefficients
\begin{eqnarray}
V_0 + V_1 + V_2 + V_4 + V_5 - U  &  \sim  &  
   O ( \Lambda_0^{1/2}  )  \nonumber      \\
V_3 - V_2    &  \sim  &    0    \nonumber          \\
V_6 - V_5    &  \sim  &   O ( \Lambda_0^{1/4} )
\label{s1}
\end{eqnarray}

In order to close this system of equations we need an
additional piece of information, that we get by estimating 
$\Pi (\theta , -\theta)$. This object stands for the scattering
with a momentum transfer connecting a point on the Fermi line
and its symmetric with respect to the $y$ axis. Therefore, it
has again the characteristic dependence $\sim O (
\Lambda_0^{1/2} / \alpha^{1/2} )$ if $\theta = 0$ or $\pi$ and it
shows a crossover to a behavior $\sim O ( \Lambda_0 )$ for other
values of $\theta $. We may approximate then
$\Pi (\theta , -\theta) \approx c + d \cos^2 (\theta )$, where 
$c \sim O ( \Lambda_0 )$ and $d \sim O (
\Lambda_0^{1/2} )$. Comparing with (\ref{exp}), this
implies that
\begin{eqnarray}
V_0 - V_1 - V_3 - V_4 - V_6 - U  &  \sim  &  
   O ( \Lambda_0 )   \nonumber     \\
2 V_1    &  \sim  &  O ( \Lambda_0^{1/2} )   \nonumber    \\
V_2 + V_3    &  \sim  &   0     \nonumber        \\
V_4   &  \sim  &   0        \nonumber        \\
V_5 + V_6   &  \sim  &   0  
\label{s2}
\end{eqnarray}
Putting together (\ref{s1}) and (\ref{s2}), we obtain $V_6 \sim
- V_5 \sim O ( \Lambda_0^{1/4} )$, $V_0 - U \sim O (
\Lambda_0^{1/4} ) $ and $V_1 \sim O ( \Lambda_0^{1/2} )$, the
rest of the coefficients being higher-order powers of 
$\Lambda_0^{1/4}  $.

Our analysis shows that there is a negative coupling at an
intermediate energy scale $\Lambda_0 $ in the expansion in
normal modes of the marginal interaction $V (\theta , \theta') $.
The negative coupling $V_5$ sets actually the leading behavior
of the coefficients $V_n$, for $n \neq 0$, as the energy cutoff
is sent to the Fermi level. The approximations we have made of
$\Pi (\theta , \theta)$ and $\Pi (\theta , -\theta)$ by
specific periodic functions do not have therefore major
influence in the determination of the attractive channel, as
long as the higher harmonics in the expression (\ref{exp})
can be considered subdominant with respect to the behavior of
the couplings $V_1$, $V_5$ and $V_6$.

We conclude that there must be a range of dopings, around the
filling level where the inflection points are evenly distributed 
in the angular variable on the Fermi line, in which a
superconducting instability opens up in the channel
corresponding to the $A_1$ representation of the lattice
symmetry group\cite{prl}. This means that the order parameter of
superconductivity has the so-called extended $s$-wave symmetry,
with nodes at the inflection points of the Fermi line.
This location of the nodes has to be shared to a certain
degree of approximation by any Fermi line with a set of
inflection points. The gap in the superconductor tends to close up
at the points where the interaction gets more repulsive. In
the model with local interaction, this happens at the points
where there is more phase space for the scattering between the
particles. The character of the bare interaction is crucial
in this consideration, since it relies on the fact that the main
effect of the anisotropy at the scale $\Lambda_0 $ comes through
diagram $(c)$ in Fig. \ref{prl}. This contribution goes in the
opposite direction to screening the bare interaction. It becomes
clear that, under a more general type of interaction, different
possibilities for the opening of an attractive channel could
arise, while the methods outlined in this Section should still
be useful to study very anisotropic Fermi lines.

In the context of the $t-t'$ Hubbard model, the situation in
which the Fermi line approaches the saddle points at $(\pi ,0)$
and $(0, \pi)$ has also great phenomenological interest. One
possibility would be to address this problem by means of a
sequence of different dopings and Fermi lines having the kind of
inflection points we have just discussed. However, this would imply
that, when approaching the saddle points, two inflection
points would become quite close near $(\pi ,0)$ and $(0, \pi)$.
It is clear that the kind of approximations we have made before
should break down at a certain point in this limit, as the
distribution of the strength of the scattering along the Fermi
line becomes quite sharp. In the limit case where the Fermi
level is at the Van Hove singularity, the distribution of the
curvature of the Fermi line becomes actually singular at the
saddle points. This suggests that a different method has to be
deviced to deal with this special situation, on which we will
elaborate in the following Section.

Anyhow we still have the alternative of approaching the saddle
points from above the Van Hove
singularity, with the kind of round shaped Fermi lines shown
in Fig. \ref{fermil}. As far as the curvature is distributed smoothly
over the Fermi line, we can approximate again the
BCS vertex by a certain number of harmonics, the
rest of them being much more irrelevant as the cutoff is
reduced.

For the round Fermi lines depicted in Fig. \ref{fermil} 
the bare vertex 
$\Pi (\theta , \theta)$ is 
$\sim O ( \Lambda_0^{1/2} / \alpha^{1/2} )$, so that it is just 
modulated by the local curvature $\alpha $ of the Fermi
line. It is a function with period equal to $\pi /2$, reaching
maxima at the angles where the curvature has the minimum value.
For convenience, we measure now angles with the origin at 
$(\pi ,\pi )$.
We may take then 
\begin{equation}
\Pi (\theta , \theta ) \approx a + b \sin^2 (2 \theta )
\end{equation}
where $a \sim b \sim O ( \Lambda_0^{1/2} )$. Comparing with the
expansion (\ref{exp}), we obtain the relations
\begin{eqnarray}
V_0 + V_1 + V_2  - U  &  \sim  &  
   O ( \Lambda_0^{1/2}  )   \nonumber     \\
V_3 - V_2    &  \sim  &   O ( \Lambda_0^{1/2} )
\label{d1}
\end{eqnarray}
Using as before the estimate $\Pi (\theta , -\theta) \approx c + d
\cos^2 (\theta )$, with $c \sim O ( \Lambda_0 )$ and $d \sim O (
\Lambda_0^{1/2} )$,
we have the additional constraints
\begin{eqnarray}
V_0 - V_1 - V_3  - U  &  \sim  &  
   O ( \Lambda_0 )  \nonumber      \\
2 V_1    &  \sim  &   O ( \Lambda_0^{1/2}  )   \nonumber     \\
V_2 + V_3    &  \sim  &   0
\label{d2}
\end{eqnarray}

Taking into account (\ref{d1}) and (\ref{d2}), we conclude that 
$V_3 \sim - V_2 \sim  O ( \Lambda_0^{1/2}  ) $ and 
$V_1 \sim  O ( \Lambda_0^{1/2}  ) $. In this case, we find 
that among the dominant contributions as $\Lambda_0 \rightarrow 0$ there is
a negative coupling, $V_2$, that corresponds to an instability
in the  $d_{x^2 - y^2}$ symmetry channel. This result concerning
the order parameter is in agreement with more detailed
calculations for Fermi lines of similar shapes\cite{zanchi}.

Thus, although we are not able to deal yet with the particular
instance in which the Fermi level sits at the Van Hove
singularity, the above discussion makes plausible that a strong
superconducting instability may develop with $d_{x^2 - y^2}$
order parameter at that special filling. The detailed study of
the electron system at the Van Hove singularity shows that there
is competition with magnetic instabilities, although for
intermediate values of $t'$ the pairing condensate develops at a
higher energy scale and superconductivity prevails\cite{japon}. 
For the $t-t'$ Hubbard model with $t' <
0$, we are led to propose a phase diagram with a superconducting
instability with $d_{x^2 - y^2}$ order parameter above the Van
Hove filling, for intermediate values of $t'$ ( above $\approx - 0.3$) . 
Below the Van
Hove singularity we have found that there is a range of dopings
where the superconducting instability has extended
$s$-wave pairing symmetry, with nodes at the location of the
inflection points.

Our description of the anisotropic Fermi lines relies on the
assumption that these do not change much upon lowering the
cutoff $\Lambda_0$. The RG approach provides a consistent
picture if only the electrons near the Fermi line are affected
by the interaction, that is, $\Lambda_0 \sim U \ll
\varepsilon_F$, where $\varepsilon_F$ is the Fermi energy.
Different superconducting instabilities are known to arise
in very dilute systems, for instance,
where this approximation is not
valid\cite{rusos}.

\section{Saddle points at the Fermi line}

In this section we analyze the instabilities of a two--dimensional
electron system whose Fermi line passes through the saddle points 
$(\pi,0)=A$ and $(0,\pi)=B$ shown in Fig. 1 . To extract the physics
of the saddle point
we will take as an example the Hubbard t--t' model where  perfect
nesting is absent.
We think that most of the features that we get are due to the
presence of the saddle
points in the Fermi line more than to the specific model.

As mentioned in the
previous section, this case can be seen as the limiting
 situation in which two inflection points merge into a saddle
point; the technique developed previously can not be directly
applied since the Fermi surface geometry becomes  singular in
this case.

The presence of  saddle points at the Fermi line
has been related to the physics of the cuprates
from the very beginning as a possible explanation for their high
$T_c$ superconductivity
\cite{hove,schulz,dz,markg,mark2,tsuei,patt,epl,ioffe,npb}.
Its interest was reinforced by the photoemission
experiments\cite{photo,gofron} showing
that the hole-doped materials tend to develop very flat bands near the
Fermi level. The subject has evolved into the so--called Van
Hove scenario that can be studied in the literature.  In
this section we will review the features of the model that   
can be extracted from a renormalization group point of view  
and derive its Kohn--Luttinger superconductivity
following  closely the scheme set in the previous  sections.

As described in section 2, the starting point of the RG
study for fermion systems is the
bare action despicted in (\ref{effact}):
\begin{eqnarray}
S  & = & \int dt d^2 p \left( i \Psi^+_\sigma ({\bf p}) \partial_t
   \Psi_\sigma ({\bf p}) - \left( \varepsilon ({\bf p}) - \varepsilon_F
  \right) \Psi^+_\sigma ({\bf p}) \Psi_\sigma ({\bf p})  \right)
                                     \nonumber  \\
 &  &  + \int dt d^2 p_1 d^2 p_2 d^2 p_3 d^2 p_4
   U ({\bf p}_1, {\bf p}_2, {\bf p}_3, {\bf p}_4) \Psi^+_\sigma ({\bf p}_1)
   \Psi^+_{\sigma'} ({\bf p}_2) \Psi_{\sigma'} ({\bf p}_4)
       \Psi_\sigma ({\bf p}_3)
  \delta ( {\bf p}_1 + {\bf p}_2 - {\bf p}_3 - {\bf p}_4 ) \;\;\;,
\nonumber
\end{eqnarray}
including the four fermion interaction term.

Of the next two steps followed in section 2, which
are the keypoints to obtain the Fermi liquid behavior, none can be
implemented in our case.  The first is associated
to the isotropy of the Fermi line and consists in the
kinematical decomposition of the momenta into parallel and
orthogonal to the Fermi line at each point. It makes the
scaling behavior of  the meassure in the action effectively one-dimensional
and is at the origin of the kinematical classification of the
marginal channels.
The second one is the
Taylor expansion of the dispersion relation $ \varepsilon ({\bf p})$
to the linear order what makes the momenta to have the same scaling
behavior as the  energy.
In the present case the dispersion relation associated to the
t--t' Hubbard model
\begin{equation}
\varepsilon({\bf k}) = -2t\;[ \cos(k_x a)+\cos(k_y a)]
-4t'\cos(k_x a)\cos(k_y a)
\;\;\;,
\label{disp}
\end{equation}
can be expanded around each of the saddle points A, B of Fig. 1
as

\begin{equation}
\varepsilon_{A,B} ( {\bf k} ) \approx \mp ( t \mp 2 t' ) k_x^2 a^2
\pm ( t \pm 2 t' ) k_y^2 a^2  \;\;\;,
\end{equation}
where  the momenta $k_x, k_y$ measure
small deviations from $A, B$.
The quadratic dependence of the dispersion relation on the
momenta induces the scaling tranformation:
$$\;\omega\rightarrow s\omega\;\;,
\;\; {\bf k}\rightarrow s^{1/2}{\bf k}\;\;\;.$$

Now it is easy to see that the same scaling law of the Fermi
fields obtained in the Fermi liquid analysis: 
 $\;\; \Psi ({\bf p}) \rightarrow
s^{-1/2}  \Psi ({\bf p})\;,$
makes the free action marginal
---since the scaling of the integration measure is the same---,
and makes the four-fermion interaction to be marginal irrespective of
the kinematics as now the delta function  scales as the
inverse of the momentum for a generic kinematics.
 
Next, also in contradistinction with 
what happens in the marginal couplings of the
Fermi liquid, the renormalization of the couplings in the Van
Hove model is nontrivial due to the logarithmic divergence  of the
density of states dictated again by the dispersion relation.
As described in section 3, the density of electron--hole pairs
at a given momentum ${\bf q}$ depends on the curvature of the Fermi
surface at the two points connected by ${\bf q}$.   

This situation is similar to the one-dimensional model 
where there are also logarithmic singularities
in the particle--hole and particle-particle propagators.
As we will now see, the similarity stops at the
classification level. The one loop analysis shows that singular 
quantum  corrections arise in the saddle point model only when
the momentum transfer in the process is of the forward (F),
exchange (E) or BCS (V) type. Moreover the different channels do not
mix at this level making the study more similar to the 
Fermi liquid case.  

In order to make clear the former statement we start by
performing the standard analysis of the Van Hove g--ology. Although 
inclusion of spin is not necessary for the description of the
Kohn--Luttinger
superconductivity, we shall take it into account to include magnetic
instabilities.

The complete classification of the marginal interactions including the two
flavors A and B follows exactly the one that occurs 
in the g-ology of one--dimensional systems \cite{lutt,libro} where the
role of the two Fermi points is here played by the two singularities.

\begin{figure}
\begin{center}
\epsfysize=12cm
\epsfbox{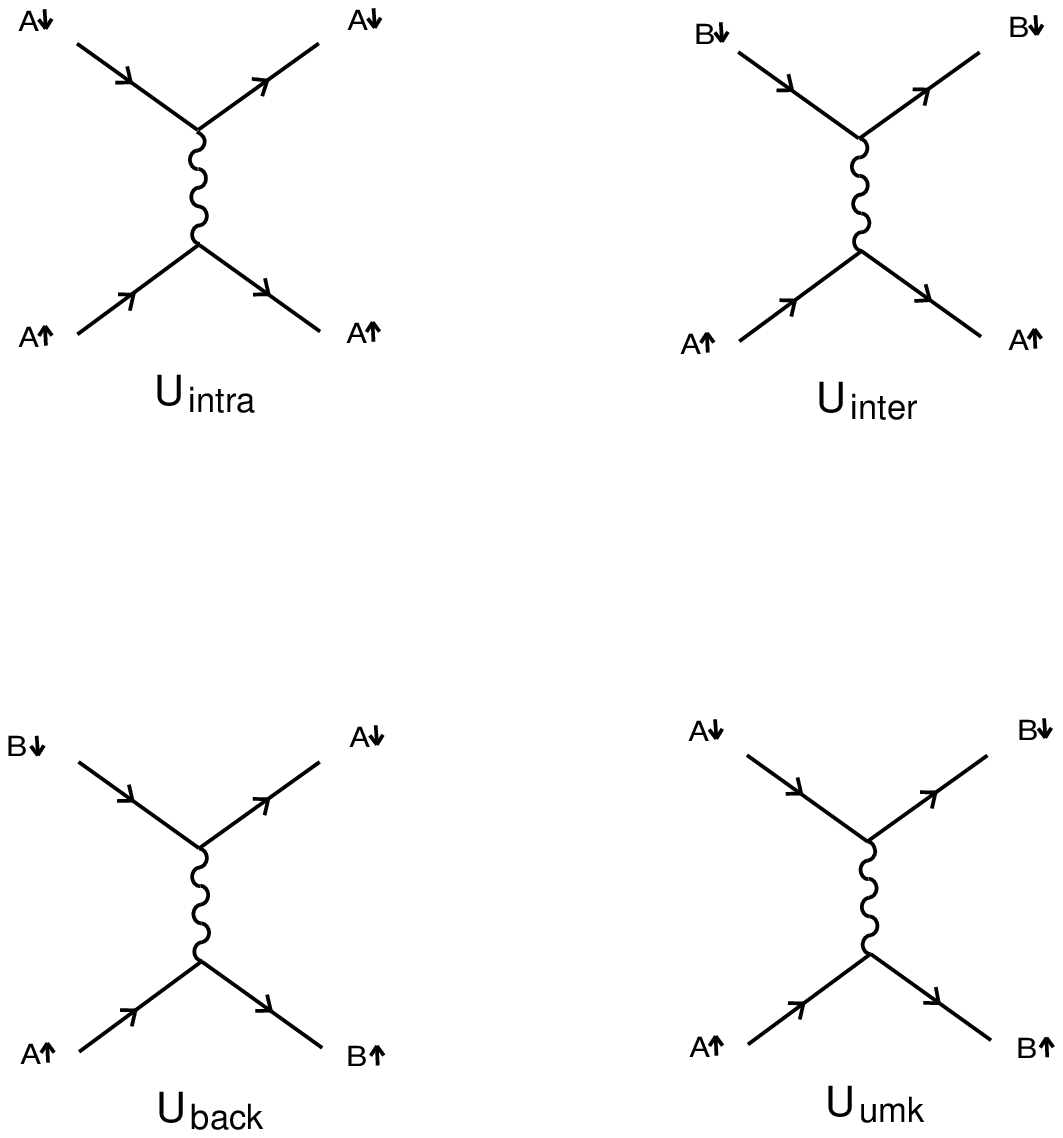}
\vspace{14pt}
\end{center}
\caption{Different interaction terms arising from the flavor indices
A and B .}  
\label{couplings}
\end{figure}

In general there are four types of interactions
that involve only low-energy modes. They are displayed in  
Fig. \ref{couplings}  where 
the interaction is represented 
by a wavy line to clarify the process it refers to. 

In the model that we
are considering, with a constant interaction potential, 
the wavy lines should be shrunk to a point giving rise
to couplings typical to the $\Phi^4$ quantum field theory.
The spin indices of the currents are opposite in
all cases  to stay as close as possible to   
the original Hubbard interaction. In this spirit, parallel
couplings or spin--flip
interactions shall not be allowed.

\begin{figure}
\begin{center}
\epsfysize=8cm
\epsfbox{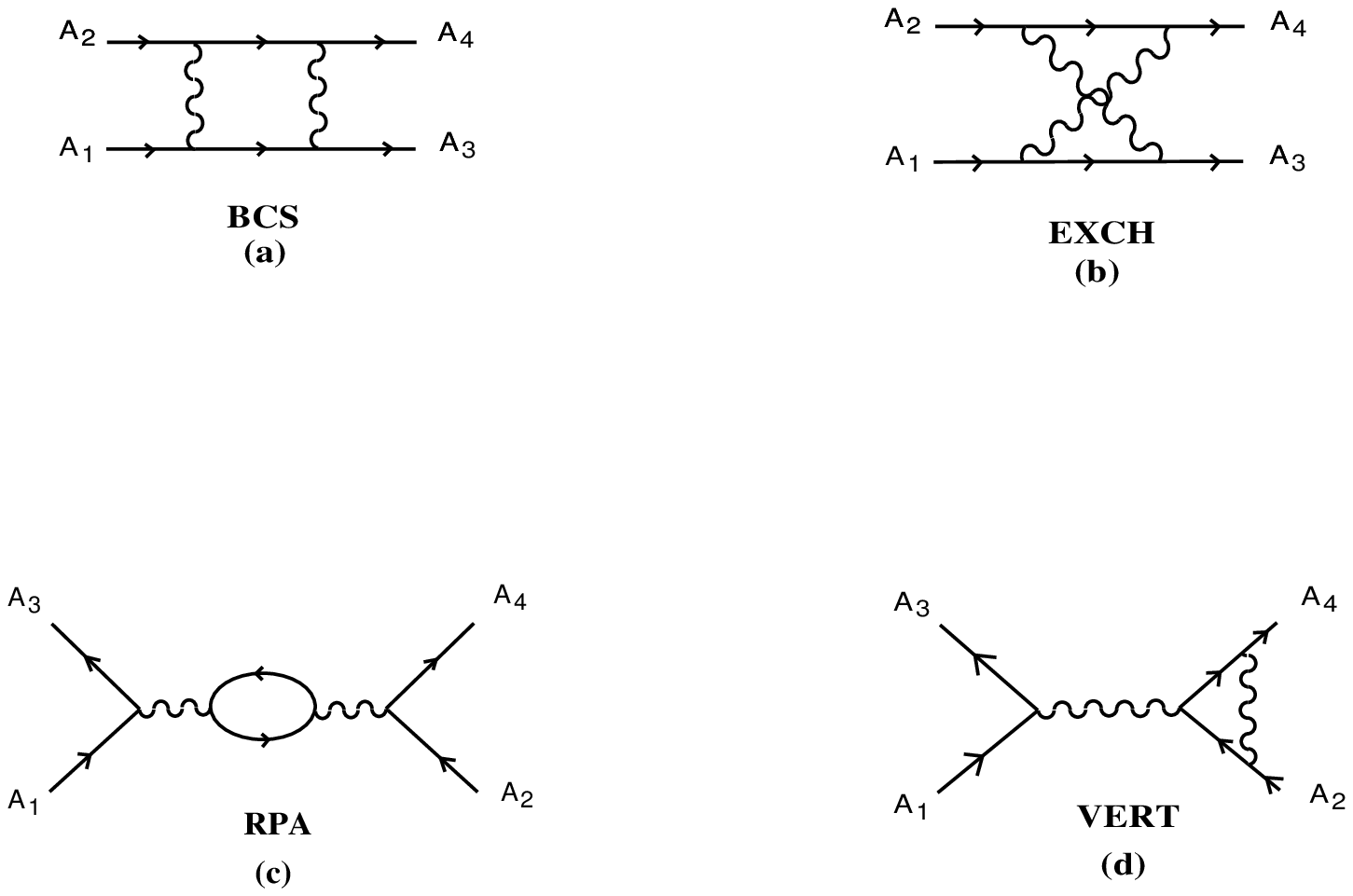}
\vspace{14 pt}
\end{center}
\caption{Diagrams contributing to the one-loop order correction
to the interaction potential.}
\label{three}   
\end{figure}

All one loop corrections to a generic four fermion coupling are
depicted in Fig. \ref{three}.

There are four types of corrections in Fig. \ref{three}: direct (BCS)
and exchange (EXCH)  interactions 
(Fig. \ref{three} (a), (b)),
particle--hole interactions called RPA in Fig. \ref{three} (c), 
and vertex corrections called VERT in Fig. \ref{three} (d).

\begin{figure}
\begin{center}
\epsfysize=9cm
\epsfbox{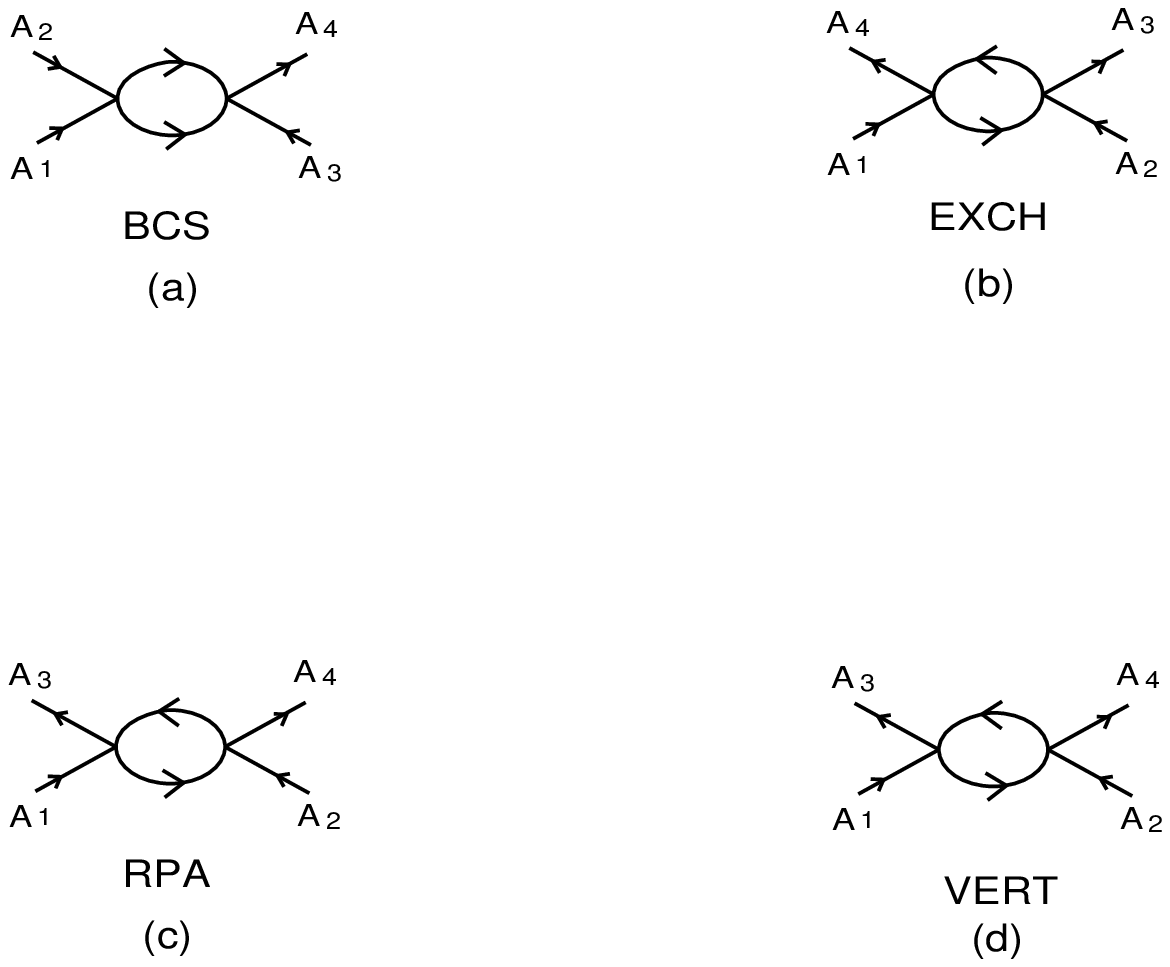}
\vspace{14 pt}
\end{center}
\caption{Diagrams of Fig. \ref{three} with the interaction
shrunk to a point.}
\label{four}
\end{figure}

Once the interactions are shrunk to a point, they turn into the 
diagrams depicted in Fig. \ref{four} which are the ones to be computed 
in the one-loop calculation. We must keep in mind that, according to
Fig. \ref{three}, the corrections induced by the BCS and RPA diagrams
(Figs. \ref{three} (a) and \ref{three} (c)) 
have a minus sign relative to the others.

All the one--loop diagrams  can be parametrized in terms of 
the  four polarizabilities involving 
particle--particle or particle--hole processes 
with the two lines corresponding to the same (zero momentum transfer)
or to different
flavors A, B (momentum transfer ${\bf Q}$). 
These have been computed in \cite{epl,npb,ferro},  their
cutoff--dependent part at energy $\omega=0$ is 

\begin{eqnarray}
\chi_{{\rm ph}}  ({\bf q}) & \sim &
\frac{c}{2 \pi^2 t} \log\vert \frac{\Lambda}{\varepsilon ({\bf q})}\vert
\nonumber  \\
\chi_{{\rm ph}} ({\bf Q}) & \sim &
\frac{c'}{2 \pi^2 t} \log\vert\frac{ \Lambda}{ta^2({\bf q}
-{\bf Q})^2}\vert
\nonumber  \\
\chi_{{\rm pp}}  ({\bf q})  & \sim & 
\frac{c}{4 \pi^2 t} \log^2\vert \frac{\Lambda}{\varepsilon ({\bf q})}\vert
\nonumber  \\
\chi_{{\rm pp}}  ({\bf Q})  & \sim &
\frac{1}{4 \pi^2 t'}\arctan (2ct'/t) 
 \log\vert\frac{ \Lambda}{ta^2({\bf q}
-{\bf Q})^2}\vert
\;\;, \label{pp}
\end{eqnarray}
where 
\begin{equation}
c \equiv 1/\sqrt{1 - 4(t'/t)^2} \;\;\;,\;\;\; c' \equiv
\log \left[ \left(1 + \sqrt{1 - 4(t'/t)^2} \right)/(2t'/t) \right]\;\;\;.
\label{ccp}
\end{equation}
Leaving aside for a moment the BCS graph, we see that all
one--loop corrections are written in terms of the particle--hole
polarizabilities (\ref{pp}) 
which diverge if the momentum transfer
is zero or ${\bf Q}$. That means that the RPA and VERT graphs
will only provide corrections to a coupling if
the kinematics of the vertex is such that 
$p_1\sim p_3\;\;,\;\;p_2\sim p_4\;\;,$ i.e. both are forward processes
in the Fermi liquid language
which will renormalize only the amplitudes having the
specified kinematics.
The  EXCH graph of Fig. \ref{three} will in turn be divergent only for the
kinematics  $p_1\sim p_4\;\;,\;\;p_2\sim p_3\;\;,$
i.e. it corresponds to an exchange process which renormalize
exchange amplitudes.

\begin{figure}
\begin{center}
\epsfysize=6cm
\epsfbox{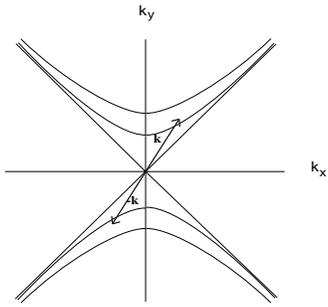}
\vspace{14pt}
\end{center}
\caption{Two energy slices used in the computation of the BCS graph.}
\label{slices}
\end{figure}

The analysis of the BCS channel 
is similar to the one done for the isotropic Fermi
line. It is easy to see that particle--particle processes will
provide a logarithmic renormalization only to those 
couplings whose kinematics is fixed to be of the BCS type. In
the differential approach that we are using,
this is best seen graphically as depicted in Fig. \ref{slices}. 

Two energy
integration  slices contributing to the computation of the
BCS graph are shown in Fig. \ref{slices}. It is clearly seeen that, unless
the total momentum of the incoming particles adds to zero, the
area of the intercept of the two bands for which two high energy
intermediate states contribute in the loop (which measures the cutoff
dependence of the diagram) is of  order $(d \Lambda)^2$.
This is different from what happens in one dimension, where the BCS
graph contributes to the renormalization of all quartic couplings.
Moreover, forward or exchange
processes do not contribute to the BCS flow.

We are then faced to
a very similar situation as in the Fermi liquid case. Instead of
talking of renormalization of couplings, we must
analyze the fate of each channel forward, exchange and BCS,
independently as they will not get mixed at the one loop level.

Let us first analyze the F channel. It is easy to see that at 
the one loop level and for the case that we have chosen to analyze of 
point like 
interactions between currents of opposite spins, this coupling does not
flow
since  we can not draw the diagrams of
the corresponding 
corrections without invoking
parallel couplings in the case of the RPA diagrams or spin flip
interactions
in the VERT diagram of \ref{three}. In the more general case in which
parallel
couplings are included from the beginning, there is an exact cancelation
between the two types of F couplings as can be seen from Fig. \ref{three}.
For each diagram of RPA type that can be drawn there exists
another of VERT type that cancels it.
This cancellation that, to our knowledge, was first noticed in the
original work of ref. \cite{kl} will occur to all orders in perturbation
theory if the given conditions hold. Any k-dependence of the 
interaction  would destroy this symmetry. In particular the very
process of renormalization necessarily induces that dependence. In the
presence of parallel couplings, the two graphs  (c) and
(d) in Fig. \ref{three}
will renormalize differently giving rise to a flow for the F
channel. We will not address this question here. A related
problem is discussed in \cite{nd}.

The flow of the exchange channel must be examined by for each of the
couplings of Fig. \ref{couplings} that will from now
on be denoted by $E$ meaning that the
momenta of the external legs are fixed to the exchange kinematics.
The RG equations are obtained by 
``opening up'' the graph and inserting the polarizabilities
in such a way that the resulting graph is of the type of Fig.
\ref{three} (b) and the vertices are made up of the tree--level 
interactions of Fig. \ref{couplings}.

\begin{figure}
\begin{center}
\epsfysize=16cm
\epsfbox{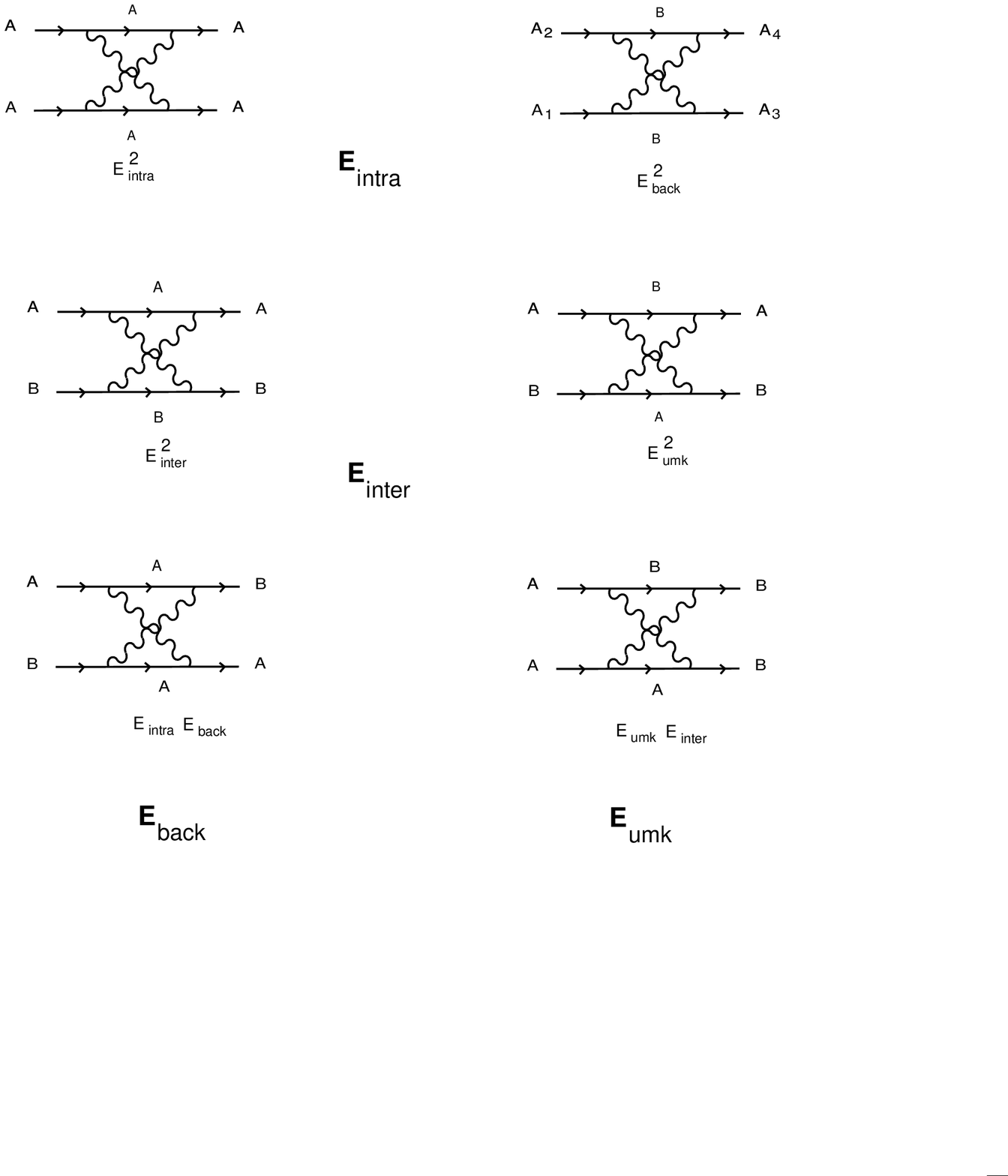}
\vspace{14pt}
\end{center}
\caption{Renormalization of the different couplings described in
the text.}
\label{corr}
\end{figure}

Let us first discuss the behavior of any coupling, say $E_{{\rm inter}}$.
$E_{{\rm inter}}$ is renormalized  by the diagrams shown
in Fig. \ref{corr}.

Adding up the one--loop correction to the bare coupling we find 
the vertex function at this order
\begin{equation}
\Gamma_{{\rm inter}}(\omega) \approx E_{{\rm inter}}+
\frac{c'}{2 \pi^2t} ( E_{{\rm inter}}^2+E_{{\rm umk}}^2 )
\log \left|\frac{\Lambda}{\omega}\right|  
\end{equation}
Following the usual RG procedure, we define the dressed
coupling constant at this level in such a way that the vertex function be 
cutoff independent. Defining $l=\log\Lambda$, we get the RG equation
\begin{equation}
\frac{\partial E_{{\rm inter}}(\Lambda)}{ \partial l}= -
\frac{c'}{2\pi^2t} ( E_{{\rm inter}}^2 + E_{{\rm umk}}^2 )\;,
\label{beta}
\end{equation}
were the polarizability involved is the interparticle polarizability
and the  sign of the beta function is negative as corresponds to the
``antiscreening'' diagram of Fig. \ref{three} (b). 
The same equation (\ref{beta}) is obtained by integrating
over a differential energy slice as we mentioned earlier.

By this method we obtain the following set of coupled differential
equations for
the $E$ couplings:
\begin{eqnarray}
\frac{\partial E_{{\rm intra}}}{\partial l}  & = & -
 \frac{1}{4\pi^2 t} c \left(E_{{\rm intra}}^2 + E_{{\rm back}}^{2}
       \right)    \nonumber   \\
\frac{\partial E_{{\rm back}}}{\partial l}  & = &  -
 \frac{1}{2\pi^2 t} c \left( E_{{\rm intra}}  E_{{\rm back}} \right)
\nonumber \\
\frac{\partial E_{{\rm inter}}}{\partial l}  & = &  -
 \frac{1}{4\pi^2 t} c' \left( E_{{\rm inter}}^2 + E_{{\rm umk}}^{2} 
\right)  \nonumber\\
\frac{\partial E_{{\rm umk}}}{\partial l}  & = &   -
 \frac{1}{2\pi^2 t} c' \left( E_{{\rm inter}} E_{{\rm umk}}
        \right)      \label{flow}
\end{eqnarray}
where $c, c'$
are the prefactors of the polarizabilities at zero and ${\bf Q}$
momentum transfer, respectively given in (\ref{ccp}).

Due to the relative plus sign of the corrections, all $E$ couplings
will grow if repulsive giving rise to instabilities in the system 
that can be identified by means of the response functions. It is
easy to see that the $E$ kinematics couples to the response
functions corresponding to the order parameters 
$$\Delta_{SDW}^x = \sum_k b^+_{ks}\sigma_{st}^x 
a_{k+Q,t}\;\;,$$
$$\Delta_{FM}^x = \sum_k a^+_{ks}\sigma_{st}^x 
a_{kt}\;\;.$$
The growth of these response functions drives the system to 
antiferromagnetic and ferromagnetic ground states respectively.
The phase diagram can be seen in \cite{japon}.

Let us now see how  the Kohn-Luttinger mechanism works in this
case. As discussed in Section 2,
superconducting instabilities will arise in the system by
means of the BCS channel whenever a negative coupling develops.
RG equations can be obtained for the BCS coupling by the same
procedure followed with the $E$ couplings. We will denote the BCS couplings 
by $V$  implying the prescribed kinematics
in the couplings of Fig. \ref{couplings} and correct them at one
loop by the BCS graphs in Fig. \ref{three}. 
As we are looking for pairing
instabilities we will only consider in Fig. \ref{three} (a)
couplings with total momentum equals zero (not ${\bf Q}$). This leaves
us with $V_{{\rm intra}}$ and $V_{{umk}}$.   
Both involve the polarizability $\chi_{pp}(0)$ 
(\ref{pp}).
We will take care of the $\log^2$ dependence of the particle--particle 
susceptibility by the method discussed in \cite{npb}. It consists in
noting that the $\log^2$  can be factorized as the usual RG log and 
another one which is due to the divergent density of states exactly at the
Fermi line and
will take a constant value when the renormalization of the chemical
potential
is taken into account. We will denote this constant value by
$K=\log(\Lambda /\mu)$.
Proceeding as before, we get for the $V$ couplings the following
set of equations: 
\begin{eqnarray}
\frac{\partial V_{{\rm intra}}}{\partial l}  & = &
 \frac{1}{4\pi^2 t} c K \left( V_{{\rm intra}}^2 + V_{{\rm umk}}^{2}
       \right)   \nonumber \\
\frac{\partial V_{{\rm umk}}}{\partial l}  & = &
 \frac{1}{2\pi^2 t} c K \left( V_{{\rm intra}}  V_{{\rm umk}} \right)
\nonumber
\end{eqnarray}
As there is no mixture of particle--hole channels, integrating this
equation
is equivalent to summing the ladder series. 

\begin{figure}
\begin{center}
\epsfysize=7cm
\epsfbox{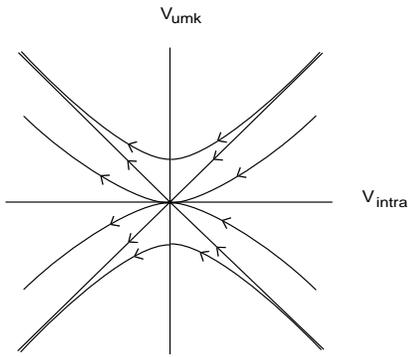}
\vspace{14pt}
\end{center}
\caption{The flow of the V couplings described in
the text.}
\label{flowf}
\end{figure}

The flow of the couplings is shown in Fig. \ref{flowf}. It is clear that if
all the couplings are set to a common value at the beginning,
say the value $U$ of the original Hubbard model, then 
$ V_{{\rm intra}} = V_{{\rm umk}} $ all along 
and they would flow to zero
along the diagonal in Fig. \ref{flowf}.

The Kohn--Luttinger mechanism will be at work if 
 at some stage of the flow we get as initial condition
$ V_{{\rm intra}} - V_{{\rm umk}} < 0 $. This difference can be
caused by the finite corrections induced in the BCS vertices 
by the particle--hole diagrams in Fig. \ref{three} at the early stage of
the flow. As discussed in \cite{shankar}, that an operator
is irrelevant means that it
will flow to zero in the course of the renormalization, but
it does not mean that it can be set to zero at the beginning
without altering the physics. In this case, the BCS couplings 
$ V_{{\rm intra}}$ and $ V_{{\rm umk}} $ will get finite
particle--hole corrections from the exchange diagram that make them differ
at
intermediate values of the cutoff providing the initial
conditions necessary for Kohn-Luttinger superconductivity. 

From equations (\ref{flow}) of the exchange couplings it is clear that
under
the initial condition that all couplings are equal to $U$,
$E_{{\rm intra}} = E_{{\rm back}}$ 
and $E_{{\rm inter}} = E_{{\rm umk}}$, all along the flow. 
As the corrections coming 
from the $E$ vertex have a positive sign,  
we see that in the region of parameter space where $c<c'$ in (\ref{ccp}),
we will soon get $ V_{{\rm intra}} < V_{{\rm umk}} $ and then
the RG flow of the BCS diagram will start with the initial condition 
necessary for  
d-wave superconductivity to set in.

\section{Conclusions}

In this paper we have analyzed the instabilities induced on a 
strongly correlated electron system in two dimensions by the  
anisotropies of its Fermi line. Although we have chosen to
exemplify the features with a Hubbard $t-t'$ model we believe
that the study done is very general and will apply for other
systems as well. Our main conclusion is that changes in the
topology of the Fermi  line of an electron system
driven by doping, pressure or
whatever means, give rise to instabilities of the system
that will drive it away from Fermi liquid behavior. In  the absence
of nesting or other accidental symmetries, the instabilities
will be predominantly superconducting.
Although this idea is not new, as it goes back
to the Kohn--Luttinger mechanism of the sixties, we have been
able to see it at work in a  microscopic model with
different kinds of anisotropies.

We have used a renormalization
group  approach which tracks the origin of the Kohn--Luttinger
mechanism on the effect that some irrelevant operators have on
marginal couplings. The scaling analysis allows us to predict, on
very general grounds, a critical dimension of 3/2 for the onset
of non--Fermi liquid behavior, a result that remains to be
tested. Inflection points, a generic feature of some Fermi
lines, give rise to a superconducting order parameter with
extended $s$--wave symmetry with nodes located at the inflection
points.

The case of extreme anisotropy of the Fermi line having
saddle points has been analyzed with the same technique,
obtaining a very rich phase diagram for the $t-t'$ Hubbard model
with $d_{x^2-y^2}$ superconductivity as the main instability in
a certain range of parameters. The same symmetry of the order
parameter is found for fillings above the saddle
points, what suggests the possibility of a continuum
transition in the overdoped regime of hole--doped cuprates.

Despite the theoretical inspiration of this paper,
the analysis performed contains some phenomenological
predictions that can be tested experimentally in compounds which
are well described by the $t-t'$ Hubbard model. In particular,
the closeness of a transition towards a $d$--wave and an extended
$s$--wave superconductor implies that, in the absence of perfect
tetragonal symmetry, a mixture of the two is likely.
This possibility has been already discussed on phenomenological
grounds\cite{ds}. Different experiments, like photoemission\cite{photo1}
or interlayer tunneling in BSCCO\cite{Li1}, can be interpreted in terms
of the coexistence of different order parameters, or the existence
of extended $s$--wave superconductivity.
Note, finally, that a recent proposal for the Fermi surface of BSCCO,
based on photoemission experiments\cite{photo2}, is close to the
case with inflection points discussed here. Such a Fermi surface leads
naturally to an extended $s$--wave order parameter in the superconducting
phase. It would be interesting to see if this prediction is confirmed by 
experiments.

\vspace{1cm}
{\em Note added } --- After this review has been published, 
the role played by the renormalized forward scattering interactions near a
Van Hove singularity has been clarified in Ref. \onlinecite{new}.
Furthermore,
two detailed RG studies have appeared\cite{halbo,hone}  
supporting the idea that $d$-wave superconductivity takes
place in the $t-t'$ repulsive Hubbard model near the Van Hove 
filling.

\newpage

\newpage

\newpage

\newpage

\newpage

\newpage

\newpage

\newpage

\newpage

\newpage


\begin{references}
\bibitem{rgkl}
An elaboration from the point of view of the renormalization group
approach can be found in Ref. \onlinecite{shankar}.
\bibitem{follow}
We follow at this point the developments in Refs. \onlinecite{shankar}
and \onlinecite{pol}.
\end{references}
\end{document}